# Estimation of Apollo lunar dust transport using optical extinction measurements

John E. LANE[1] and Philip T. METZGER[2,3]

[1]Easi-ESC, GMRO Lab, Kennedy Space Center, FL, USA

e-mail: john.e.lane@nasa.gov (corresponding author)

[2]NASA Granular Mechanics and Regolith Operations, KSC, FL, USA

[3]Florida Space Institute, University of Central Florida, Orlando, FL, USA

e-mail: Philip.Metzger@ucf.edu

## Abstract

A technique to estimate mass erosion rate of surface soil during landing of the Apollo Lunar Module (LM) and total mass ejected due to the rocket plume interaction is proposed and tested. The erosion rate is proportional to the product of the second moment of the lofted particle size distribution $N(D)$, and third moment of the normalized soil size distribution $S(D)$, divided by the integral of $S(D) \cdot D^2/v(D)$, where $D$ is particle diameter and $v(D)$ is the vertical component of particle velocity. The second moment of $N(D)$ is estimated by optical extinction analysis of the Apollo cockpit video. Because of the similarity between mass erosion rate of soil as measured by optical extinction and rainfall rate as measured by radar reflectivity, traditional NWS radar/rainfall correlation methodology can be applied to the lunar soil case where various $S(D)$ models are assumed corresponding to specific lunar sites.

# GLOSSARY OF SYMBOLS

| Symbol | Description | Standard Units | *si* Units |
|---|---|---|---|
| $N(D)$ | lofted particle size distribution | $\mu m^{-1}\ m^{-3}$ | $m^{-1}\ m^{-3}$ |
| $D$ | particle diameter | $\mu m$ | $m$ |
| $S(D)$ | normalized soil size distribution, empirical fit | $\mu m^{-1}$ | $m^{-1}$ |
| $S_1(D)$ | component of $S(D)$ fit | - | - |
| $S_2(D)$ | component of $S(D)$ fit | - | - |
| $w(D)$ | component of $S(D)$ fit | - | - |
| $D_1 \ldots D_3$, $B_1 \ldots B_3$ | fitting constants in particle size fraction model $S(D)$ | $\mu m$<br>dimensionless | $m$<br>dimensionless |
| $v(D)$ | vertical component of particle velocity | $m\ s^{-1}$ | $m\ s^{-1}$ |
| $R$ | rainfall rate | $mm\ h^{-1}$ | $m\ s^{-1}$ |
| $Z$ | radar reflectivity | $mm^6\ m^{-3}$ | $m^3$ |
| $a$ and $b$ | parameters of Z-R, Eq. (1) | - | - |
| $A$ and $B$ | parameters of $\alpha$-$\dot{m}$, Eq. (3) | - | - |
| $\dot{m}$ | lunar soil erosion rate | $kg\ s^{-1}\ m^{-2}$ | $kg\ s^{-1}\ m^{-2}$ |
| $\Gamma(D)$ | Gamma Function of $D$ | - | - |
| $\lambda$ | wavelength of light | $nm$ | $m$ |
| $\rho_L$ | bulk density of lunar soil | $g\ cm^{-3}$ | $kg\ m^{-3}$ |
| $M_x$ | $x$th moment of size distribution | $mm^x\ m^{-3}$ | $m^{x-3}$ |
| $\varphi$ | geometry factor in Eq. (4) | dimensionless | dimensionless |
| $s(f)$ | particle shape factor | dimensionless | dimensionless |
| $\alpha$ | optical extinction factor | $\mu m^2\ m^{-3}$ | $m^{-1}$ |
| $x$ | particle size parameter | dimensionless | dimensionless |
| $r_a, r_b$ | short and long radius of particle ellipsoid, respectively. | $\mu m$ | $m$ |
| $f$ | particle shape factor, $r_a / r_b$ | dimensionless | dimensionless |
| $Q_e$ | scattering efficiency factor for extinction | dimensionless | dimensionless |
| $n$ | refractive index | dimensionless | dimensionless |
| $a_0(t)$ | ideal radius of surface erosion as a function of time $t$ | $m$ | $m$ |
| $h(t)$ | nozzle opening distance from surface as a function of time $t$ | $m$ | $m$ |
| $m_T$ | total mass rejected (total mass displaced) | MT = 1000 kg | kg |

| Symbol | Description | Units | Units |
|---|---|---|---|
| $u(h, D)$ | CFD based particle maximum speed model | m s$^{-1}$ | m s$^{-1}$ |
| $b_0 \ldots b_4$, $c_0, c_1$ | empirical fitting constants in CFD particle speed model | - | - |
| $\eta$ | mean CFD particle speed compared to maximum. | dimensionless | dimensionless |
| $V(r)$ | particle speed fit, $r$ is the trajectory starting point (radial distance from nozzle center) | m s$^{-1}$ | m s$^{-1}$ |
| $R_1, R_2$ | min and max CFD domain distance for determining $\eta$ | m | m |
| $v_0, v_1, \xi$ | fitting constants in $V(r)$ fit | m s$^{-1}$, m s$^{-1}$, m | m s$^{-1}$, m s$^{-1}$, m |
| $\sigma$ | CFD derived shear stress | N m$^{-2}$ | N m$^{-2}$ |
| $\mu$ | plume gas dynamic viscosity | kg m$^{-1}$ s$^{-1}$ | kg m$^{-1}$ s$^{-1}$ |
| $T(r, z)$ | CFD gas temperature, $r$ from nozzle center, $z$ above surface | K | K |
| $v_r$ | radial component of the plume gas velocity | m s$^{-1}$ | m s$^{-1}$ |
| $T_0, T_C, \mu_0$ | constants in Sutherland's formula for shear stress | - | - |
| $\bar{\sigma}(h)$ | CFD shear stress at surface, averaged over radial distance, versus engine height | N m$^{-2}$ | N m$^{-2}$ |
| $\sigma_0, \Gamma$ | shear stress fitting parameters, Eq. (13) | N m$^{-2}$, m$^{-1}$ | N m$^{-2}$, m$^{-1}$ |
| $m_0, \Lambda$ | Erosion rate fitting parameters, Eq. (13) | kg, m$^{-1}$ | kg, m$^{-1}$ |
| $c, \sigma_c, \beta$ | parameters in theoretical shear stress model | m s$^{-1}$, N m$^{-2}$ | m s$^{-1}$, N m$^{-2}$, dimensionless |

## 1. INTRODUCTION

Previous work has focused on particle trajectory analysis and computational fluid dynamics (CFD) simulations of rocket plume interactions with the lunar surface (Morris, *et al*. 2011, Immer *et al*. 2011a, Lane *et al*. 2010). An important component that may be missing from a pure trajectory simulation is intensity of dust dispersal, or more precisely, the "soil mass erosion rate. In previous work, an erosion rate was estimated from the optical extinction of a few ideal image features (Immer *et al*. 2011b, Metzger *et al*. 2010). The drawback of those methods is the limited data that is available for analysis, usually only a few frames from an entire landing video. Recently a new approach was taken, following the methodology used by the National Weather Service (NWS) in measuring rainfall intensity (hydrometeor mass accumulation and intensity rate) using Weather Surveillance Radar (Wexler and Atlas 1963, Rosenfeld *et al.* 1993).

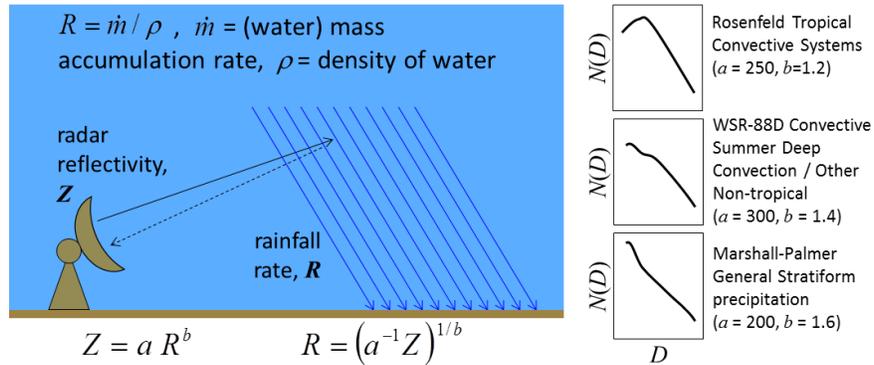

Fig. 1. Radar measurement of rainfall rate is affected by the local DSD, *N*(*D*).

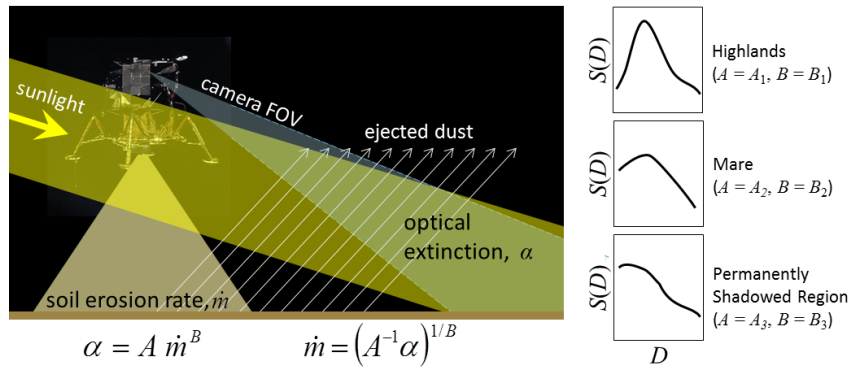

Fig. 2. Optical measurement of dust erosion rate is affected by the local soil size distribution *S*(*D*). Note that *A* and *B* are found from *S*(*D*) associated with soil properties at a specific region on the lunar surface, as well as the properties of the gas exiting the rocket nozzle used to generate *v*(*D*), based on the specific engine design.

　　The key to this approach is to assume a particle size distribution. Even though it may appear to be a risky assumption, this methodology has been in operational use by the NWS since the advent of weather radar. Part of this methodology relies on a strategy of substituting appropriate drop size distribution (DSD) functions for specific meteorological regimes (tropical, continental, etc.). This is then similar to substituting different soil size distributions for various areas on the lunar surface where surface operations and landings are planned, such as highlands, mare, and permanently shadowed craters (highland and mare samples were returned by Apollo missions, but return of permanently shadowed craters samples will be a goal of some future missions). Once the erosion rate is determined for a specific rocket engine, it can then be used to correlate those predictions with CFD simulations which also predict surface shear stress due to the rocket plume interaction with the surface (Metzger *et al*. 2011). Note that CFD simulations are needed to navigate from engine design specifications to predictions of surface shear stress predictions and dispersed particle velocities.

　　Another result of this study is that it becomes obvious that optical extinction due to scattering of light from hydrometeors can be used to estimate rainfall rate, just as microwave radar may be used to measure soil erosion rate (Lane and Metzger 2014b). This resemblance is a consequence of the similarity of size range of the particle distributions of hydrometeors and lunar soil and the fact that the index of refraction which determines the details of electromagnetic scattering is similar. Before this equivalence is taken too far however, it must be recognized that 10 cm weather radar does not detect fog size particles, which are comparable in size to the smallest lunar dust particles that may contribute to soil erosion. Therefore, to measure dust particles effectively, millimeter wave radar would need to be utilized to correctly quantify this analogy.

## 2. MASS EROSION RATE

Measuring lunar soil erosion rate $\dot{m}$ from optical extinction $\alpha$ during a rocket landing is analogous to measuring terrestrial rainfall rate $R$ using NWS radar reflectivity $Z$. The similarity in both cases is the dependence on some knowledge of the particle size distribution functions, $N(D)$. For weather radar, the quantity of interest $R$ is usually reported in mm h$^{-1}$. For soil erosion, the quantity of interest $\dot{m}$ is measured in units of kg s$^{-1}$ m$^{-2}$. Note that the product of rainfall rate and water density also has units of kg s$^{-1}$ m$^{-2}$.

The weather radar Z-R relation is a simple power-law and correlates measured reflectivity $Z$ and rainfall rate $R$ using two parameters, $a$ and $b$ (see Figure 1). Radar reflectivity $Z$ is the sixth moment of the DSD. By parameterizing and fitting $N(D)$ so that its integral is a Gamma Function, $\Gamma(D)$, equations for $Z$ and $R$ can be combined to form a power law of the form:

$$Z = aR^b \qquad . \qquad (1)$$

where $a$ and $b$ are related to the parameters that describe the drop size distribution and drop terminal velocity functions. For operational use, these parameters are determined empirically from weather radar data and networks of rain gauges. Note that the size distribution curves shown in the right hand side of Figures 1 and 2 are for illustration only and do not depict actual size distributions.

Similarly, the soil erosion rate $\dot{m}$ is found by integrating the product of the particle mass and velocity times the size distribution $N(D)$. The optical extinction is found from the second moment of the size distribution (Atlas 1953). Shipley *et al.* (1974) demonstrated empirically that optical extinction has a power law relationship with the rainfall rate and therefore with the suspended hydrometeor mass when the drop size parameter $x \equiv \pi D / \lambda >> 1$, where $\lambda$ is the illumination wavelength. This condition applies to particles such as lunar dust when $D >> \lambda / \pi$.

At this point the similarity diverges since the size distribution $N(D)$ is the lofted distribution of particles and in the lunar soil case is an unknown. Rainfall DSDs can be measured directly using ground based or aircraft based disdrometers. However, the normalized soil size distribution $S(D)$ can be measured (using lunar samples returned to Earth) and is related to $N(D)$ and the CFD simulated particle velocity $v(D)$ according to:

$$S(D) = \frac{v(D) N(D)}{\int_0^\infty v(D) N(D) \, dD} \qquad . \qquad (2)$$

Now the relationship between optical extinction $\alpha$ and mass erosion rate $\dot{m}$ can be approximated by a power-law, analogous to the hydrometeor case of Equation (1):

$$\alpha = A\dot{m}^B \qquad , \qquad (3)$$

where the parameters $A$ and $B$ are found using the soil size distribution $S(D)$ associated with soil properties at a specific region on the lunar surface, as well as the properties of the gas exiting the rocket nozzle used to generate $v(D)$, based on the specific engine design.

Calculation of soil mass erosion from optical extinction can be approximated using the following equation:

$$\dot{m}(t) = \frac{\rho_L \pi \varphi}{6 \, s(f)} M_2(t) \, \frac{\int_0^\infty S(D) D^3 \, dD}{\int_0^\infty S(D) \left(D^2 / v(D)\right) dD} \qquad , \qquad (4)$$

where $M_2(t)$ is the second moment of the lofted size distribution, which is indirectly measured by video camera analysis. A DSD moment is defined as:

$$M_x = \int_0^\infty D^x N(D) dD \ . \tag{5}$$

The bulk density of lunar soil $\rho_L$ is approximated as 3100 kg m$^{-3}$. S(*D*) is the normalized soil size distribution as measured by an image analysis based particle size analyzer, particle sieves, or some equivalent method. The parameter $\varphi$ is a geometry factor accounting for the divergence of the dust ejecta, spreading radially outwards from the nozzle centerline. It can be shown that $\varphi \approx 2$ using a simple model of the dust ejection pattern (see Appendix A). For the case of zero divergence, such as terrestrial rainfall, $\varphi = 1$.

The particle shape factor *s*(*f*) is a function of the particle aspect ratio $f = r_b/r_a$ where $r_a$ is the short radius and $r_b$ is the long radius. In the idealized case of the *prolate spheroid*, all quantities involving *D* are computed as usual with a diameter $D = (r_a^2 r_b)^{1/3} = r_a f^{1/3}$. The shape factor for particles modeled as a prolate spheroid with aspect ratio *f* is, $s(f) = (\pi + 2(f-1))/(\pi f^{2/3})$ (see Appendix B).

The optical extinction factor $\alpha$ is related to the second moment as $\alpha = (\pi/4) Q_e M_2$ where $Q_e$ is the scattering efficiency factor for extinction (Berg *et al*. 2011). In general, $Q_e$ is a function of the size parameter *x* and for a narrowband fixed spectrum of light, it can be approximated as a function of only particle size *D*. As shown by Hulst (1957), the minimum size factor *x* that determines the boundary between classical and Mie scattering is a function of refractive index of the scattering particle. The larger the refractive index *n*, the smaller the threshold value of *x*. In the case of hydrometeor scatterers, $n = 1.33$, so classical scattering with $Q_e = 2$ applies to drop sizes for $x > 6$, and for a spectrum centered about green light ($\lambda_G = 532$ nm), $D > 6\lambda_G/\pi$. Therefore visible light can be used to measure the extinction factor of hydrometeors greater than 1 µm using a constant $Q_e = 2$. In this case, the second moment of the size distribution is related to the extinction factor $\alpha$ by a factor of $2/\pi$. The same is true in the case of lunar dust particles, with the exception of a larger index of refraction. In the lunar dust case with $n = 1.75$ (ignoring the small imaginary component), Figure 24 of Hulst (1957) shows $x > 3$, so that $Q_e = 2$ can be used down to a particle size of 0.5 µm, again using the visible light spectrum centered about $\lambda_G$.

The soil size distribution *S*(*D*) for Apollo 11 sample 10084 and Apollo 17 sample 70051 are shown to have a peak around 0.030 µm (Metzger *et al*. 2010). This would seem to violate the assumption $x > 3$ and $Q_e = 2$. However, it is not the peak of *S*(*D*) that is relevant to the issue of $Q_e$, it is the moments of the soil size distributions as expressed by the numerator and denominator of the right side of Equation (4). It can be shown that the peaks occur around $D = 30$ µm, well within the assumption that $x > 3$ and $Q_e = 2$.

### 2.1. Extinction Factor and Effective Radius of Erosion

Figure 3 shows the time dependence of the LM cockpit video frame sequence for the final 60 s of descent and for 60 s after landing and engine cutoff. The dotted line represents the video frame number at 12 fps rate, while the line with open circles plots the LM height according to the voice recording of altitude radar callout. The upper black line is the histogram average of each video frame, while the lower green line is the corresponding standard deviation. The histogram plots of Figures 18 and 19 (Appendix C) show a bimodal characteristic. Crater shadows and a low sun illumination angle produce the low end histogram peak (dark shadows) while the high end peak is due to everything else. Dust in the image has the effect of forcing these two peaks together since dark shadows become lighter due to the backscatter of light from the dust, and light areas become darker due to extinction of the reflected light from the surface. This process is revealed primarily by a reduction of the standard deviation, caused by the peaks converging as the dust cloud density increases.

Region A of Figure 3 can be used as a reference since it is a clear image of a typical surface scene. Other regions are described in Table 1.

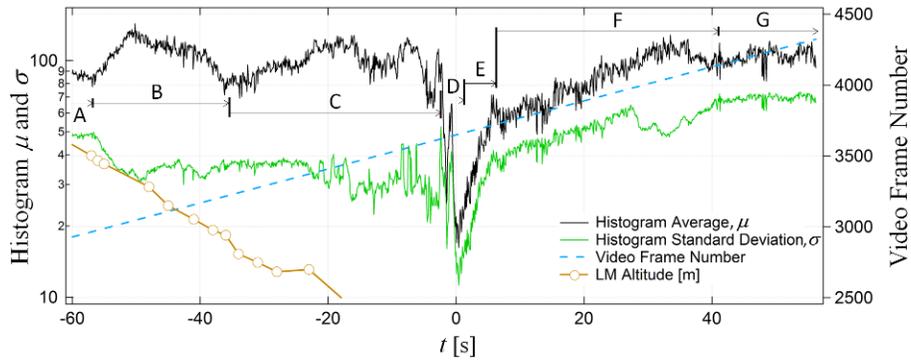

Fig. 3. Histogram parameters for the final 60 s of descent and for 60 s landing and after engine shutoff.

Table 1. Description of regions in Figure 3.

**A**: *Frame 2962*. Images in this region can be used as a histogram reference. The variation of the histogram average and standard deviation is minimal during this segment of the video. Frames showing unusually large dark craters need to be excluded from the reference baseline.

**B**: *Frame 3111*. Images in this region of the video are good candidates for the histogram matching method (HMM), described in Appendix C.

**C**: *Frame 3404*. Images in this region are not suitable for the HMM. Manual selection of extinction parameters is done in this region by trial-and-error, comparing the dust treated image visually.

**D**: *Frame 3647*. Images in this region experience a total blackout due to LM shadows on the top of the dust cloud. As the dust cloud settles the shadow quickly disappears. This region is not useable for extracting extinction information due to the effect of the LM shadows.

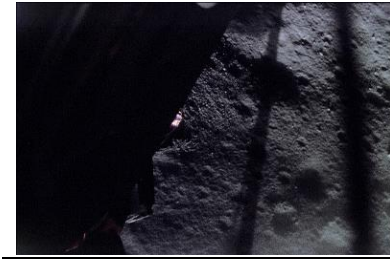

**E**: *Frame 3713*. As the dust cloud height decreases, the LM shadows fade. A smaller amount of dust persists for at least another 30 s.

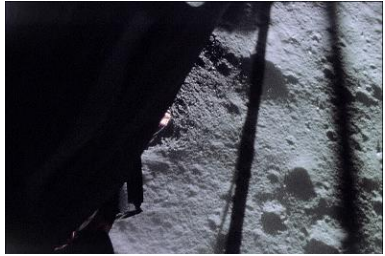

**F**: *Frame 4004*. The dust in this region is clearing at a slow rate, indicative of levitation due to effects such as electrostatic repulsion or escaping rocket exhaust gas previously forced into the regolith. The mechanism of the dust levitation is an area of current research.

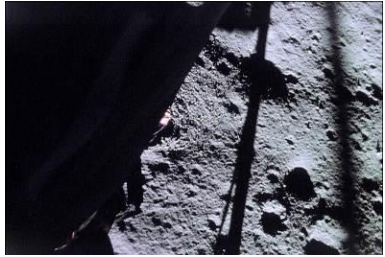

**G**: *Frame 4311*. The dust has cleared in this region. Variations of the histogram average or standard deviation are due to camera noise and/or noise introduced during the image digitization process.

Reiterating from the last section, $\alpha = \pi Q_e M_2 / 4$, where $Q_e$ is the scattering efficiency, assumed to be equal to 2. The second moment in Equation (4) is then the product of the measured extinction factor and $2/\pi$. The extinction factor $\alpha$ can be estimated from the Apollo videos by adding dust to a clear reference image (before dust appears) and comparing to the dusty image of interest. By matching histograms of the two images, the extinction factor can be estimated. The details of the histogram matching method (HMM) (Lane *et al.* 2014a) are discussed in Appendix C. HMM is more than just adjusting contrast and brightness of the images. The dust erosion angle ($\theta \approx 3°$ for Apollo 12) and radius of erosion is used to apply different amounts of contrast-brightness equalization to strips across the image, rotated to align with the horizon (and the dust cloud top). By iteratively adjusting the optical extinction factor $\alpha$ and radius of erosion $a_0$ and comparing the histogram averages and standard deviations, a best fit $a_0$ and $\alpha$ are found. It may be feasible to estimate a dust erosion angle $\theta$ using HMM, an area of possible future work.

The total mass rejected (total mass displaced) is Equation (4) integrated over vehicle descent time and over the area where soil is eroded:

$$m_T = 2\pi \int_{-\infty}^{0} \int_{0}^{a_0(t)} \dot{m}(t)\, r\, dr\, dt$$
$$= \pi \int_{-\infty}^{0} a_0^2(t)\, \dot{m}(t)\, dt \qquad (6)$$

where $a_0(t)$ is the radius on the surface, referenced to the engine nozzle centerline, where erosion is taking place. The assumption inherent in Equation (6) is that erosion is uniform over a circle of radius $a_0(t)$ and zero outside of that circle. An estimate of $a_0(t)$ for the Apollo 12 LM is shown in Figure 4, which is an output of the HMM algorithm. Note that in this and all previous discussions of $S(D)$ and its moments, it has been assumed that the particle size distribution is homogeneous over the extent of meas-urement, i.e., within a circle of radius $a_0(t)$, and all temporal effects due to engine LM altitude and thrust occur instantaneously over this spatial extent.

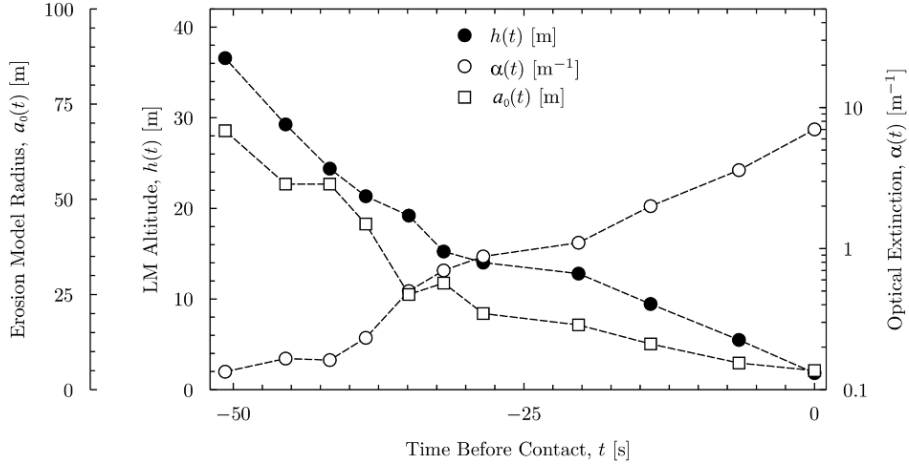

Fig. 4. Apollo 12 optical extinction estimate using histogram matching method. The time intervals correspond to the voice callouts of LM pilot, Alan Bean.

The erosion rate should actually vary with the state of the gas flowing across the soil, including its shear stress, rarefaction, and turbulence; it should also vary with saltation, including the downward flux of larger particles that are too heavy to be carried away by the gas as well as smaller particles that are scattered back down from the entrained cloud via particle collisions. Examination of the sandblasting effects on Surveyor III have shown that the downward flux of scattered particles is significant (Immer *et al.* 2011a) and discrete element computer simulations show the important but largely unexplored role of mid-flight particle scattering in enhancing erosion rate (Berger *et al.* 2013). Influence of the gas upon erosion rate should be greatest in an annular region around the vehicle (Roberts 1963) while the influence of saltation may be greater in another annulus with larger radius since particles travel downrange before striking the surface. Thus the net erosion rate may be somewhat more uniform and spread over a broader region than if gas effects alone are considered. The details of erosion physics -- especially in lunar rocket exhaust conditions -- are not yet well understood, so a constant erosion over a finite area assumed, as it is the simplest model and therefore a sensible first step.

## 2.2. Particle Velocity Function

Single particle trajectory modeling, based on CFD simulations of the Apollo LM engine and the lunar environment (Lane *et al.* 2010), yield a particle velocity function which can be described by the empirical fit shown in Equation (7), analogous to the hydrometeor terminal velocity formulas used in meteorology:

$$u(h,D) = \begin{cases} 0 & \log D + c_1 \log h > c_0 \\ \dfrac{10^{(b_1 + b_2 \log h)\tanh(b_3 + b_4 \log h)}}{b_0 \left(h^{9/20} + D^{1/2}\right)} & otherwise \end{cases}, \quad (7)$$

where $b_0 = 0.2964$, $b_1 = -0.225$, $b_2 = 0.1954$, $b_3 = 5$, $b_4 = 4.343$, $c_0 = 2.212$, and $c_1 = 3.53$ ($h$ and $D$ in meters). Equation (7) is plotted in Figure 5. For every point in $\{h, D\}$ space, a distribution of particle velocities is computed from the particle trajectory code using an equivalent Monte Carlo distribution of initial particle trajectory starting points, height above the surface and horizontal distance from the engine nozzle centerline.

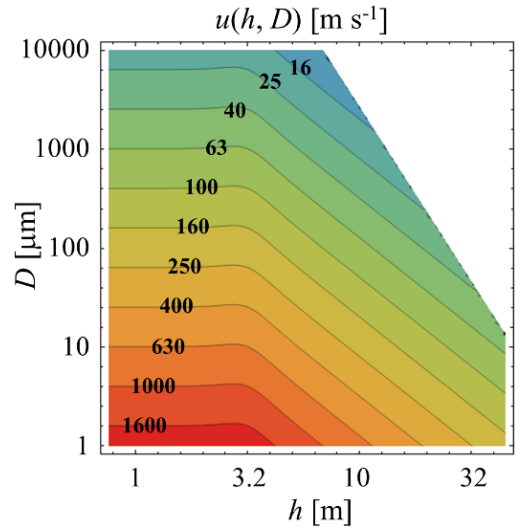

Fig. 5. Maximum particle velocity, $u(h, D)$ m s$^{-1}$.

The fit given by of Equation (7) and Figure 5 represents a maximum value of particle velocities, where particles originate near the outside rim of the rocket nozzle. The area in the upper left of Figure 5 represents the region of $\{h, D\}$ where particles do not lift from the surface due to an insufficient lifting force. Since $u(h, D)$ is a maximum velocity, the velocity in Equation (4) at each time step is replaced with a reduced value of Equation (7), $v(D) = \eta \sin\theta \, u(h, D)$, where $\theta$ is the plume propelled dust angle, equal to approximately 3 degrees (Immer *et al*. 2011b), where $\eta \leq 1$.

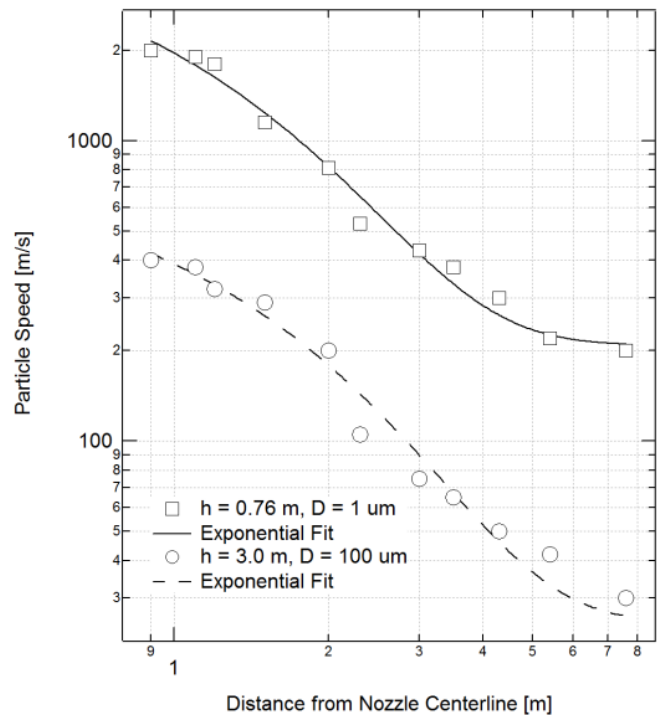

Fig. 6. Particle speed as a function of its radial starting distance from the nozzle center of the Apollo LM descent engine, for two example values of $h$ and $D$.

Because of the distribution of particle velocities for a given $h$ and $D$, the maximum values originate near the engine nozzle. Since the erosion area within $r < a_0$ (where $r$ is the radial distance from the engine nozzle centerline) is much greater for slower velocities, then it reasonable to expect $\eta$ to be much smaller than 1. The curve fits in Figure 6 are of the form $V(r) = v_0 + v_1 \exp(-r/\xi)$, which to first order are assumed to be independent of $h$ and $D$. The parameter $\eta$ can be estimated from the weighted area integral of the particle speed profile (see Figure 6) as a function of $r$:

$$\eta = \frac{2\pi \int_{R_1}^{R_2} V(r)\, r\, dr}{2\pi V(R_1) \int_{R_1}^{R_2} r\, dr} , \qquad (8)$$

where $R_1 = 0.9$ m and $R_2 = 8.0$ m are limits of the CFD particle trajectory (CFD-PT) simulations. Note that the data points along the horizontal axis are the specific values of $r$ used in the CFD-PT simulations. Performing this integral for the two curve fits in Figure 6 results in $\eta = 0.147$ for the upper curve and $\eta = 0.130$ for the lower curve. Since the value of $\eta$ does not change significantly for $D$ or $h$, a value of $\eta = 1/8$ is used as an approximation throughout the descent and erosion analysis.

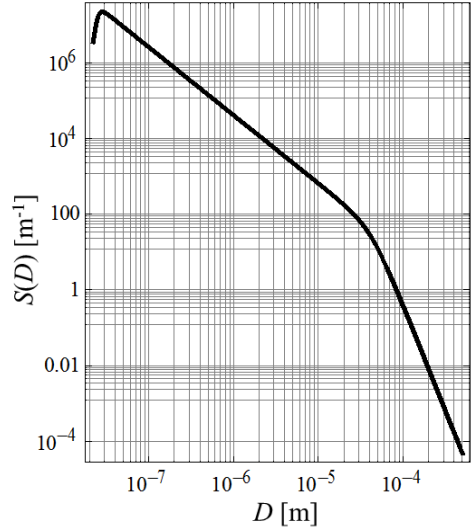

Fig. 7. Lunar regolith particle size fraction $S(D)$, estimated for Apollo 12 site using fit from Equation (9), based on samples 10084 (Apollo 11) and 70051 (Apollo 17).

## 2.3. Particle Size Distribution

The particle size fraction of the lunar soil at the Apollo 12 site is modeled as a combination of two power law functions by fitting Apollo 11 and 17 soil sample data, as well as JSC-1a simulant (Metzger *et al.* 2010):

$$S(D) = \frac{\left(w(D)S_1(D) + S_2(D)\right)^{-1}}{\int_0^\infty \left(w(D)S_1(D) + S_2(D)\right)^{-1} dD} \quad [\text{m}^{-1}] , \qquad (9)$$

where $S_1(D) = (D/D_1)^{B_1}$, $S_2(D) = (D/D_2)^{B_2}$, and $w(D) = (D_3/D)^{B_3} + 1$. The fitting constants in Equation (9) are: $D_1 = 4.090 \times 10^{-7}$ m, $B_1 = 1.8$, $D_2 = 9.507 \times 10^{-6}$ m, $B_2 = 5.6$, $D_3 = 2.5 \times 10^{-8}$ m, and $B_3 = 18$. Note that all units are kept in *si* units even though the numbers are more aesthetically pleasing in micrometers. The reason for doing this is to minimize confusion in the integrals involving $S(D)$.

### 2.4. A Justification for Equation (4)

Rather than derive Equation (4) directly, it can be worked in reverse to yield a familiar result in the meteorological case. If this is shown to be true for hydrometeors, then it follows that a particle distribution composed of granular material and dust should follow similar rules of behavior under similar forces.

Rainfall rate $R$ m s$^{-1}$ is equal to $\dot{m}/\rho$, where $\rho$ is the density of water. With this substitution, and substitution of Equation (2) for $S(D)$, Equation (4) becomes:

$$R(t) = \frac{\pi\varphi}{6} M_2(t) \frac{\int_0^\infty v(D)\, N(D)\, D^3\, dD}{\int_0^\infty v(D)\, N(D)\left(D^2/v(D)\right) dD} \quad . \tag{10}$$

$$= \frac{\pi\varphi}{6} \int_0^\infty v(D)\, N(D)\, D^3\, dD$$

For $\varphi = 1$, Equation (10) is a familiar result for computing rainfall rate in terms of the drop size distribution.

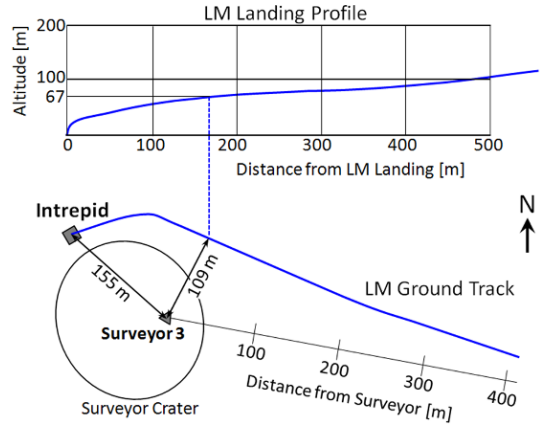

Fig. 8. Landing profile of Apollo 12 LM, Intrepid, showing Surveyor 3 landing site.

### 3. INTREPID EROSION ANALYSIS

Equation (4) computes mass erosion at each time step, with the data from Figures 4, 5, and 7, using $\eta = 1/8$, $\varphi = 2$, and $f = 1$. Table 2 summarizes these results. The sum of the eroded soil at each time step in the right column of Table 2 yields the total mass eroded. As can be seen from this table, the majority of the mass weighted erosion takes place in the last 20 s. The total eroded mass, using the parameter values chosen, equals 2594 kg. Table 3 compares the present result with previous work.

The total regolith transported from its initial resting position by erosion induced by the Apollo LM rocket engine, is estimated by integrating the mass erosion rates from Table 2 over surface area and time. Note that $t = 0$ is the surface *contact time* when the LM is approximately 1.5 m above the surface and the descent engine is turned off. Based on fall time in lunar gravity, the LM continues to descend for up to an additional 1.3 s. The index $k$ corresponding to entries in Table 2 ascend from bottom to top. The total eroded mass can be approximated by linear interpolation of $\dot{m}_k$ and $a_{0k}$ at each $k$th point:

$$m_T = \pi \int_{-10}^{0} \dot{m}(t)\, a_0^2(t)\, dt \approx \sum_{k=-10}^{0} \Delta m_k$$

$$\approx \frac{\pi}{12} \sum_{k=-10}^{0} \left( (3\dot{m}_k + \dot{m}_{k+1}) a_{0k}^2 + 2(\dot{m}_k + \dot{m}_{k+1}) a_{0k} a_{0k+1} + (\dot{m}_k + 3\dot{m}_{k+1}) a_{0k+1}^2 \right)(t_{k+1} - t_k) \quad (11)$$

The computed value of $m_T \approx 2600$ kg, and as shown in Table 3, is well within the range of other estimates of Apollo 12 total mass erosion.

Table 2. The *k*th time step corresponds to the cockpit voice recording of altitude.

| $t_k$ [s] | $h_k$ [m] | $M_{2k}$ [m$^{-1}$] | $a_{0k}$ [m] | $\dot{m}_k$ [kg s$^{-1}$ m$^{-2}$] | $\Delta m_k$ [kg], from Eq. (11) |
|---|---|---|---|---|---|
| 0 | 1.83 | 8.91 | 5.0 | 1.26 | 644.426 |
| -6.5 | 5.49 | 4.58 | 7.0 | 0.395 | 462.347 |
| -14.1 | 9.45 | 2.55 | 12.0 | 0.133 | 373.456 |
| -20.3 | 12.8 | 1.40 | 17.0 | 0.0542 | 403.772 |
| -28.5 | 14.0 | 1.12 | 20.0 | 0.0392 | 167.545 |
| -31.9 | 15.2 | 0.891 | 28.0 | 0.0282 | 208.245 |
| -34.9 | 19.2 | 0.637 | 25.0 | 0.0140 | 101.815 |
| -38.6 | 21.3 | 0.297 | 43.5 | 0.00521 | 96.013 |
| -41.7 | 24.4 | 0.206 | 54.0 | 0.00260 | 90.4892 |
| -45.5 | 29.3 | 0.211 | 54.0 | 0.00165 | 78.4725 |
| -50.7 | 36.6 | 0.171 | 68.0 | 0.000734 | 55.4542 |

Table 3. Comparison of total mass erosion estimates for Apollo landings.

| Reference | Apollo Mission | Total Mass Erosion [kg] |
|---|---|---|
| Scott (1975)[*] | Apollo 12 | 4500 to 6400 |
| Metzger *et al.* (2008) | Apollo 12 | 2400 |
| Metzger *et al.* (2010) | Apollo Average | 1200 |
| *Present work:* $m_T = \sum_{k=-10}^{0} \Delta m_k$ | Apollo 12 | 2600 |

[*]from Metzger *et al.* (2011) - interpretations of data reported by Scott (1975).

The LM landing profile shown in Figure 8 can be compared to Table 2 and Figure 3. The appearance of dust and erosion begins at an altitude of about 40 m, which is approximately 65 m from the landing site. Figure 9 shows a view of the landing site towards the south with Surveyor Crater to the left and possible surface scouring due to plume interaction, just to the left of the engine nozzle. The area of possible soil removal is shown as a discolored region, slightly browner than the greyer regolith surrounding it. Using simple image scaling, it is possible to roughly estimate the crater contour that is highlighted by the discoloration.

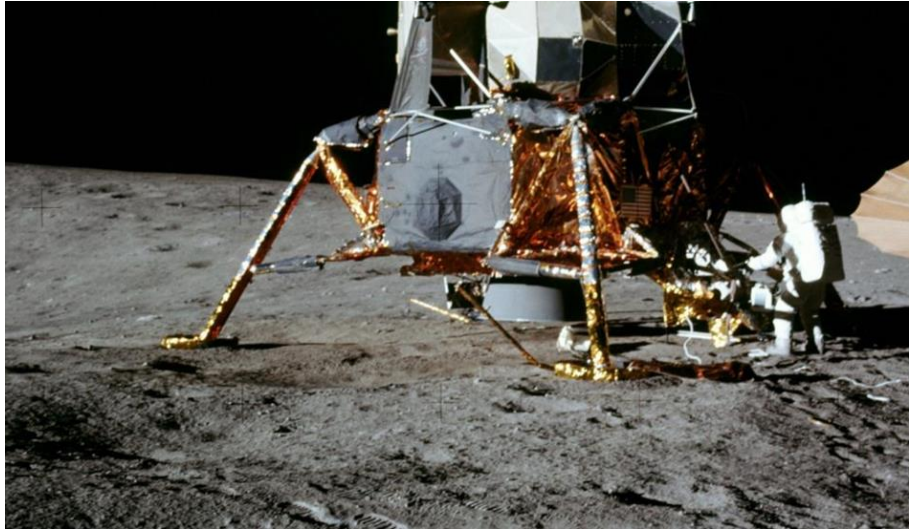

Fig. 9. View towards the south with Surveyor Crater to the left and possible scouring crater to the left of the engine nozzle (note area of discoloration).

The top of Figure 10 shows the results of a crude photogrammetry analysis of the scouring depth. The offset of the deep part of the crater is unusual and may indicate a burst of thrust just before touch-down, or a pre-existing depression in the surface. The lack of a large dug-out directly beneath the engine nozzle could be explained by the combination of a small horizontal velocity, a slight LM tilt, and engine shutoff at 1.5 m altitude (point of "contact"). For comparison, the crater from the Table 2 data is shown in in the bottom of Figure 10. The erosion picture from the optical extinction model is much shallower and greater in extent than the photogrammetry derived crater analysis.

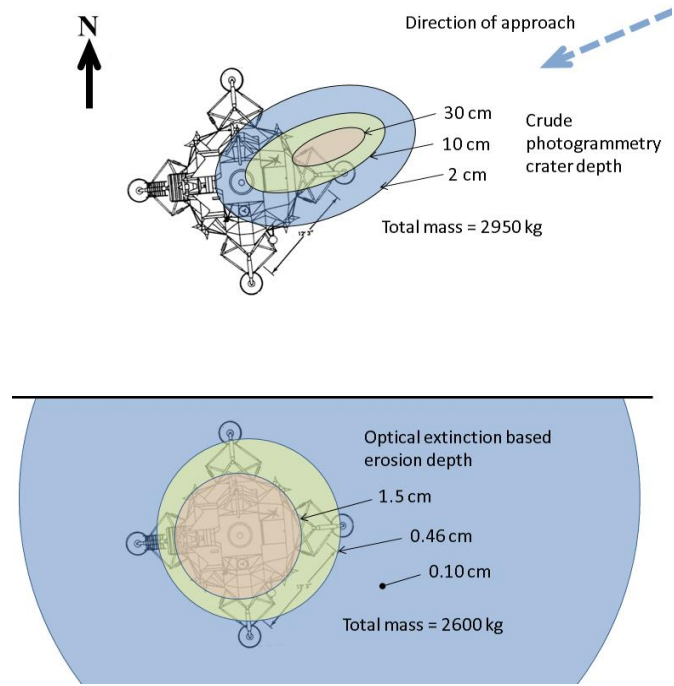

Fig. 10. Top: Crater profile based on crude photogrammetric measurements of the Intrepid landing site. Bottom: Contour map of erosion based on mass erosion from optical extinction measurements.

The size and depth of the predicted erosion depth (bottom of Figure 10) is too small to measure photographically (maximum depth is less than 2 cm). It must be pointed out that the cratering diagrammed in the top of Figure 10 is at best a hazardous guess since the previous elevation of the soil is unknown, and it is extremely difficult to get a quantifiable measurement of depth, even though multiple high resolution photographs are publically available at NASA Apollo mission archives. If the actual soil was eroded in the same fashion as the model suggests, that erosion might only be visible from a bird's eye view which might appear as a change in brightness of the surface in a large radius around the landing sit. This effect, known as *blast zone brightening*, has been observed and as yet lacks a concrete explanation (Clegg et al. 2014).

Another outcome of this study is that optical extinction due to scattering of light from hydrometeors can be used to estimate rainfall rate (Lane and Metzger 2014b, Atlas 1953), just as microwave radar may be used to measure soil erosion rate. This connection is a consequence of the similarity of size range of the particle distributions of hydrometeors and lunar soil and the fact that the index of refraction which determines the details of electromagnetic scattering is similar. Before this equivalence is taken too far however, it must be recognized that 10 cm weather radar does not detect fog size particles, which are comparable in size to the smallest lunar dust particles that may contribute to soil erosion. Therefore, to measure dust particles effectively, millimeter wave radar would need to be utilized to correctly complete this analogy.

## 4. EROSION RATE VS. SHEAR STRESS

Shear stress $\sigma$ is computed from the CFD output as a post process and is equal to the product of the dynamic viscosity $\mu$ and the vertical gradient of the radial component of the plume gas velocity $v_r$. This can be expressed in terms of gas temperature using Sutherland's formula (Smits and Dussauge 2006):

$$\begin{aligned} \sigma &= \mu \nabla_z v_r \\ &= \mu_0 \frac{T_0 + T_C}{(T(r,z) + T_C)} \cdot \frac{T(r,z)}{T_0} \cdot \frac{\partial v_r}{\partial z} \end{aligned}, \qquad (12)$$

where $\mu_0 = 1.83 \times 10^{-5}$ kg m$^{-1}$ s$^{-1}$, $T_0 = 291.2$ K, $T_C = 120$ K, and $T(r, z)$ is the gas temperature at a distance $r$ from the nozzle centerline and a distance $z$ above the surface. With this definition, shear stress has units N m$^{-2}$. Figure 11 shows the average shear stress computed by Equation (12) for four engine heights above the surface (open squares), by averaging the shear stress over the radial distance, similar to the particle velocity averaging of Equation (8):

$$\bar{\sigma}(h) = \frac{2\pi \int_0^{a_0(h)} \sigma(r,h)\, r\, dr}{\pi\, a_0^2(h)} \qquad . \qquad (13)$$

The dotted line in Figure 11 is an exponential fit of these data points, as a function of LM height $h(t)$. The mass erosion rate $\dot{m}(h(t))$ from Table 2, as defined by the model of Equation (4) (open circles), is also an average value over the area of constant radius $a_0(h(t))$. By fitting both the average shear stress to Equation (13) and average erosion rate to exponentials as a function of $h(t)$ to Equation (4), the erosion rate can be expressed as a function of shear stress by eliminating $h(t)$:

$$\begin{aligned} \sigma(t) &= \sigma_0 e^{-\Gamma h(t)} \\ \dot{m}(t) &= m_0 e^{-\Lambda h(t)} \\ \dot{m}(t) &= m_0 \left(\sigma(t)/\sigma_0\right)^{\Lambda/\Gamma} \end{aligned}, \qquad (13)$$

where $m_0 = 2.20$, $\sigma_0 = 6.21$, $\Gamma = 0.123$, and $\Lambda = 0.309$. The final relationship, based on this data, is:

$$\dot{m}(t) = 0.0222 \ \sigma^{2.52} \ \text{kg s}^{-1} \text{ m}^{-2} \quad . \tag{15}$$

## 5. SUMMARY

A method for estimating lunar soil erosion rate due to plume impingement of the Apollo 12 Lunar Module *Intrepid* during its descent to the lunar surface has been presented. The observables are optical extinction and particle size distributions of soil samples returned from the lunar surface. The optical extinction is measured between the camera mounted inside of the cockpit window and the lunar surface during landing. CFD analysis of the Apollo LM descent engine, as well as particle trajectory analysis based on the CFD simulations, provides the remainder of the necessary data.

The CFD simulations provide a key piece of information: the velocity profile of particles as a function of starting distance from the engine nozzle, size of the particle, and height of the lander from the surface. Note that there may at times be some confusion as to what is defined as height above the surface: camera, engine nozzle, or landing pads. In this paper, height $h(t)$ when used in a quantifiable analysis, is defined as the height of the engine nozzle opening to the surface, which is generally half a meter or less after landing. At other times height may refer to altitude of the landing pads.

Taking an approach similar to the problem of estimating rainfall rate from weather radar, Equation (4) was presented as the solution to the problem of estimating soil erosion rate from optical extinction measurements (see Figure 12). In both cases, the particle velocities must be known, as well as the particle size distributions. Quantifying the particle velocity function is in some sense the most difficult part of the problem for both lunar plume observations and weather radar estimation of rainfall. For this reason, the velocity function is likely the source of greatest error. More advanced coupled gas-particle flow simulations should be able to provide an improved velocity function.

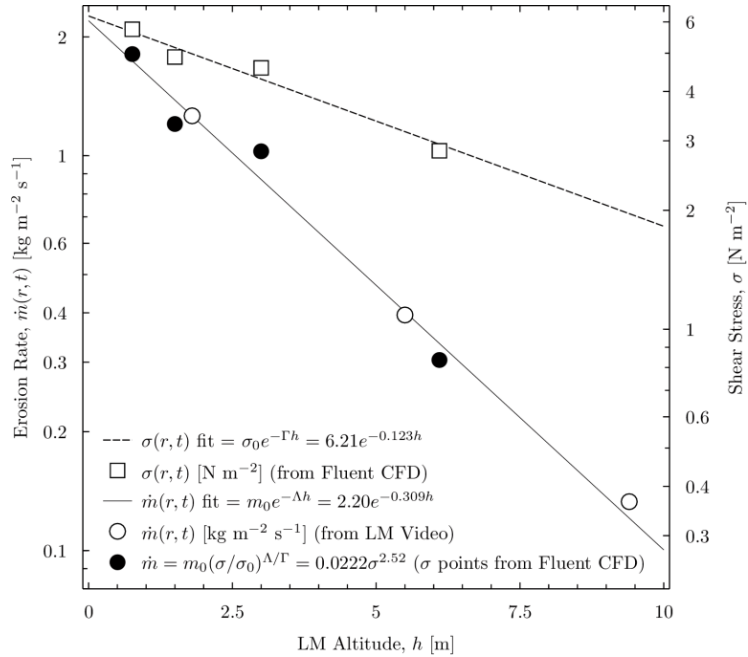

Fig. 11. Averaged shear stress and mass erosion rate as a function of *h(t)*.

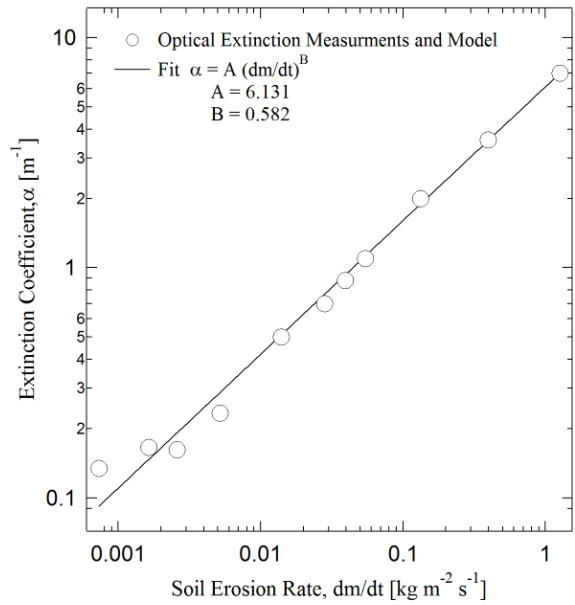

Fig. 12. Optical extinction $\alpha$ vs. soil mass erosion rate $\dot{m}$, showing power law fit.

## 5.1. Estimating Mass Erosion Rate from Optical Extinction

Equation (3) is used to estimate mass erosion rate from measurements of optical extinction once $A$ and $B$ are known. The $A$ and $B$ parameters should correspond to a particular engine design with a corresponding total vehicle mass $M$ and surface gravity $g$, which implies a nominal thrust $T = M g$ for a slow descent or hovering. The $A$ and $B$ also correspond to a specific soil type characterized by a size distribution $S(D)$.

The following summarizes the key points in determining $A$ and $B$ for the first time:

1. For various values of measured extinction factor $\alpha$, corresponding values of mass erosion rate are computed using Equation (4). These point pairs can be plotted on a log-log graph with a straight line fit to the scatter plot, providing the $A$ and $B$ parameters, as demonstrated in Figure 12. A method to determine $\alpha$ from descent videos using histogram matching has been described, as a special case.

2. The velocity $v(D)$ in Equation (4) is computed from a CFD based empirical function $u(h, D)$, such as the velocity model given by Equation (7). Then $v(D) = \eta \sin\theta\, u(h, D)$ where $\theta$ is the angle of the dust sheet relative to horizontal.

3. $\eta$ is part of the velocity model and corresponds to the ratio of mean velocity to the maximum velocity. In the example given by Equation (8), $\eta \approx 1/8$. It is assumed that $\eta$ is a constant of the engine design and has no dependencies on $S(D)$.

4. The symbol $\varphi$ in Equation (4) is a geometrical value relating the dispersion of the dust to view angle along the optical extinction path. In Appendix A, this is shown to be a constant $\approx 2$. Based on the arguments given in Appendix A, $\varphi$ should not vary much from this approximate value of 2 under varying conditions of engine design or soil type.

5. $s(f)$ in Equation (4) is an extinction shape factor which, according to the arguments given in Appendix B, can be approximated by a sphere with $s(f) = 1$, with less than 10% error for an ellipsoid when $f < 3$.

## 5.2. Image Analysis and Optical Extinction

The details of the optical extinction model and data analysis of the Apollo LM video have been deferred to Appendix C in order to avoid obscuring the details and significance of Equation (4) in estimating soil erosion rates. The optical extinction model described in Appendix C assumes from the start that the erosion rate is uniform over an area defined by radius $a_0$, which is a function of lander height. The primary influence of optical extinction as measured by a reduction of brightness of the surface and increase in brightness of the dust cloud, is the spatial dust geometry. In this model, erosion is uniform over $\pi a_0^2$ but diverges radially from all points on the surface within $r \leq a_0$. No erosion occurs for $r > a_0$. The underlying assumption of HMM is that by matching the average and standard deviation of the histogram of a reference image to the histogram of a processed image (modified by brightness and contrast equalization), the optical extinction factor can be deduced. Even though the results seem promising, the accuracy of the HMM output parameters have not been quantified. This is a possible area of future work.

### 5.3. Erosion Rate as a Function of Shear Stress

A relationship between soil erosion rate and shear stress as computed from specific engine design characteristics is highly desirable. The value of this relationship is that the total plume/erosion effects of engine design on any surface can be predicted. To this end, an empirical relationship was established between shear stress as determined by CFD simulation and erosion rate estimated by optical extinction measurements for the case of the Apollo 12 LM. The extension of these results to other engine designs for landing on any celestial body lacking an atmosphere, such as the Earth's moon or asteroids, can be used within the limits of this analysis. The result of the predicted Apollo 12 $\dot{m}(\sigma)$ relation, as shown by Equation (15), shows an approximate 5/2 power dependence of erosion rate on shear stress. Since the value of the exponent is one of the two parameters in the relationship, it is subject to sensitivity of the data measurement and analysis, as well as model assumptions. It is in fact not terribly difficult to force a linear relation (as previously believed) by substituting different values of the optical extinction data that are within credible limits of measurement error.

Equation (15) is an empirical relation for mass erosion rate $\dot{m}$ as a function of shear stress $\sigma$. Previous work concluded that the relationship should be a linear one of the form:

$$\dot{m}(t) = c^{-1}(\sigma - \sigma_c)^\beta \quad , \tag{16}$$

where $\sigma_c$ is the threshold shear stress associated with a saltation velocity threshold. The inverse proportionality constant $c$ has units of velocity. Equation (16) with $\beta = 1$ is the form predicted by Roberts (1963). Also, experiments at KSC revealed that mass erosion was proportional to the dynamic pressure of the jet leaving the pipe, i.e., $\rho v^2$, times the area of the pipe. That is also equivalent to the total thrust. It is also equal to momentum flux, which agrees with Roberts that erosion is a momentum-driven process, not an energy-driven process. According to Roberts' plume analysis theory, shear stress everywhere on the surface is proportional to thrust of the rocket, indirectly implying that the relationship is linear ($\beta = 1$). Haehnel and Dade (2008) conducted experiments where they directly measured shear stress and erosion rate locally everywhere on the surface. Erosion rate and shear stress were found to be linearly related through a global pair of constants, $\sigma_c$ and $\beta$, with $\beta = 1$.

However, complexities of the lunar case that the above three efforts do not account for include: saltation due to particles scattering out of the cloud back down to the surface; rarefaction effects; and turbulence effects, which are different in rarefied or transitional flow than in continuum flow and have never been adequately studied. Turbulence is not modeled in the existing rarefied/transitional gas flow codes. Therefore it is not unreasonable that in the lunar environment the actual value of $\beta$ may be a non-integer, as indicated by the result shown in Equation (15).

Acknowledgments. We gratefully acknowledge support from NASA's Lunar Advanced Science and Exploration Research (LASER) program, grant NNH10ZDA001N.

# APPENDIX A

## Erosion model geometry

To quantify the effect of erosion flux divergence, shown as grey arrows originating from the surface under the rocket plume in Figure 13, Equation (4) ($\varphi = 1$ describes the non-divergent case) can be applied to a small differential of erosion $\dot{m}_{jk}$. The camera image is then affected by the optical extinction occurring over a small distance $l_{jk}$ along ray $j$ due to $\dot{m}_{jk}$:

$$M_{2_{jk}} = \Phi_k^{-1} \dot{m}_{jk} \quad , \tag{A-1}$$

where $M_{2jk}$ is the optical extinction described by the second moment of the size distribution at path differential $l_{jk}$ along ray $j$; $\Phi_k$ is the collection of all other terms on the right hand side of Equation (4), characterized by velocity distribution modeled of Equation (7).

The *first model assumption* is:

$$\dot{m}_{jk} = \frac{\rho_k}{r_j} \dot{m}_k \quad , \tag{A-2}$$

which approximates the flux divergence as constrained to the shallow grey conical surface shown as arrows in Figure 13. The total optical extinction along ray $j$ is then:

$$M_{2_j} = \sum_k \frac{l_{jk}}{L_j} M_{2_{jk}} \quad . \tag{A-3}$$

Substituting Equations (A-1) and (A-2) into (A-3), and letting $l_{jk} = L_j/n$:

$$M_{2_j} = \frac{1}{n} \sum_k \Phi_k^{-1} \frac{\rho_k}{r_j} \dot{m}_k \quad . \tag{A-4}$$

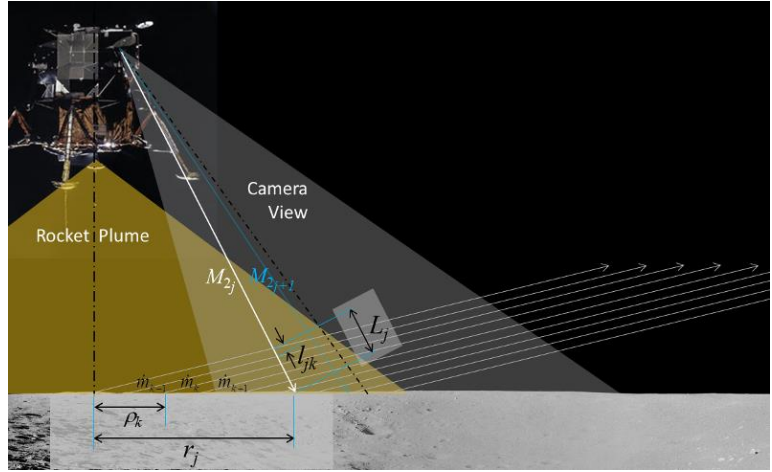

Fig. 13. Schematic diagram of erosion model geometry, leading to $\varphi = 2$ in Equation (4).

The *second model assumption* is to set all $\dot{m}_k$ equal, corresponding to constant erosion over radius $a_0$. The *third model assumption* is to set all $\Phi_k$ equal, which is in the spirit of Equation (8) where the radial dependence of particle velocity is modeled as a constant. Then Equation (A-4) becomes:

$$M_{2_j} = \frac{1}{n} \Phi_0^{-1} \dot{m}_0 \sum_k \frac{\rho_k}{r_j} \quad . \tag{A-5}$$

The radial distance $\rho_k$ in Equation (A-5) can be replaced by $k\, r_j/n$:

$$M_{2_j} = \Phi_0^{-1} \dot{m}_0 \frac{1}{n^2} \sum_k k \quad . \tag{A-6}$$

$$\lim_{n \to \infty} M_{2_j} \to \frac{1}{2} \Phi_0^{-1} \dot{m}_0$$

Comparing Equation (A-6) to Equation (4), $\varphi = 2$. Note that this is not likely a fundamental physical principle, but is more than likely a consequence of this simple model and its set of assumptions.

## APPENDIX B

### Calculation of particle shape factor for spheroid

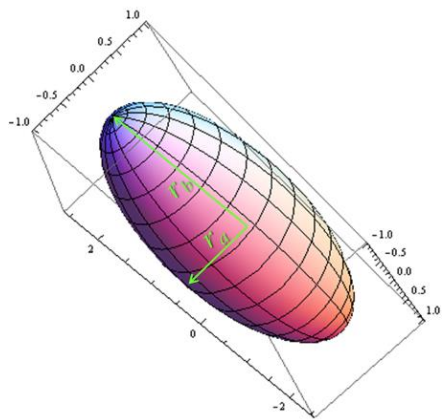

Fig. 14. Prolate spheroid with aspect ratio $f = r_b / r_a = 2.5$.

A rudimentary particle shape model, one level of improvement over a spherical particle, is the spheroid, described by aspect ratio $f = r_b / r_a$. If $f > 1$, the particle is a *prolate spheroid*. If $f < 1$, the particle is an *oblate spheroid*. And of course when $f = 1$ it is a sphere. The volume weighted diameter is $D = 2 r_a f^{1/3}$. Figure 14 shows a prolate spheroid with $f = 2.5$. The surface of the spheroid can be represented by a Cartesian vector P, which is a function of parametric angles $u$ and $v$ (similar to spherical coordinate angles $\theta$ and $\phi$):

$$\mathbf{P} = r_a \cdot \begin{pmatrix} \cos u \sin v \\ \sin u \sin v \\ f \cos v \end{pmatrix} \quad \begin{array}{c} 0 \le u < 2\pi \\ 0 \le v < \pi \end{array} \tag{B-1}$$

The average area $\overline{A}$ of a randomly oriented spheroid can be found by integrating over all values of the randomly projected major axis $r_b$, where the projection is a sinusoidal function with limits between $r_a$ and $r_b$:

$$\overline{A} = r_a^2 \int_0^\pi \left(1 + (f-1)\sin\theta\right) d\theta$$
$$= (\pi + 2(f-1)) r_a^2$$
(B-2)

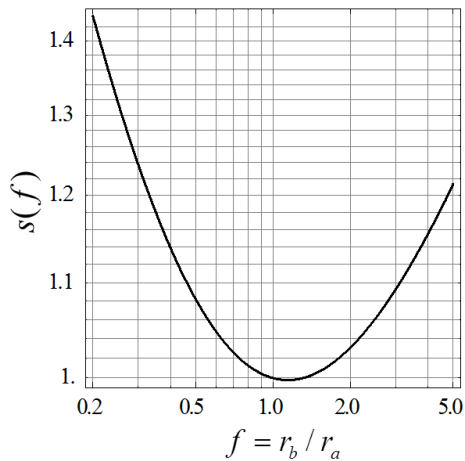

Fig. 15. Shape factor $s(f)$ used in Equation (4), described by Equation (B-3).

The shape factor $s(f)$ can then be equated to $\overline{A}$, normalized by the volume weighted cross-section of the spheroid:

$$s(f) = \frac{\overline{A}}{\pi (r_a f^{1/3})^2}$$
$$= \frac{\pi + 2(f-1)}{\pi f^{2/3}}$$
(B-3)

This result is valid for both the prolate and oblate cases. Figure 15 is a plot of $s(f)$ for $f$ ranging from 0.2 to 5. Note that in the case of $f = 2.5$, $s(f) = 1.06$, which will decrease the erosion rate of Equation (4) by approximately 6%.

# APPENDIX C

## Histogram matching method

Characterizing dust plumes on the moon's surface during a rocket landing is imperative to the success of future operations on the moon or any other celestial body with a dusty or soil surface (including cold surfaces covered by frozen gas ice crystals, such as the moons of the outer planets). The most practical method of characterizing the dust clouds is to analyze video or still camera images of the dust illuminated by the sun or on-board light sources (such as lasers). The method described below was used to characterize the dust plumes from the Apollo 12 landing.

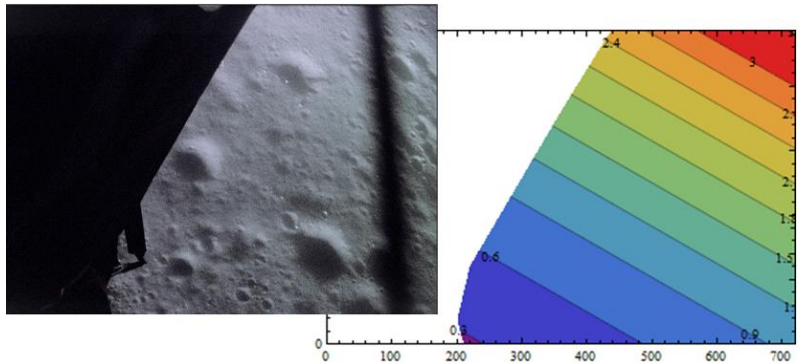

Fig. 16. Dust depth model: (left) video camera frame (F3077), with LM altitude $h = 34$ m; (right) effective camera dust length $x$ with radius $a_0 = 46$ m.

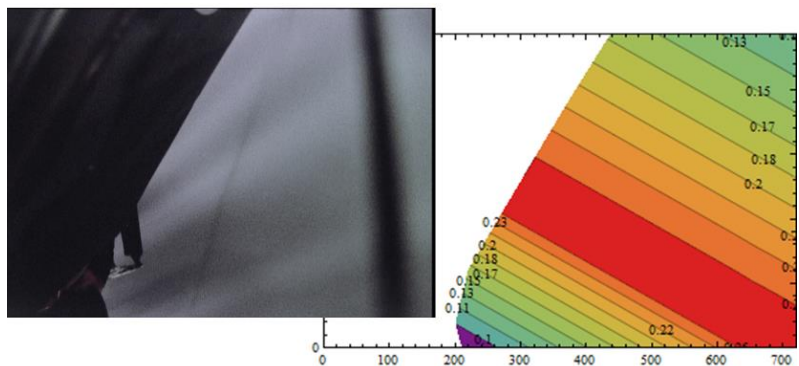

Fig. 17. Dust depth model: (left) video camera frame (F3543), with LM altitude $h = 11$ m; (right) effective dust depth with radius $a_0 = 6.5$ m.

In this context, the *histogram matching method* (HMM) is an image processing technique for determining dust optical density in Apollo landing videos. The software implementation of HMM creates a greyscale image histogram and calculates the histogram mean and standard deviation, which is then used to match dusty and clear images for the purpose of estimating an effective optical density and optical extinction factor $\alpha$. A dust thickness model, based on the tilt of the camera and increasing height of the dust layer towards the top of the image, is used to account for the distance light travels through the dust.

Previous methods relied on comparing specific features in clear vs. dusty images, which severely limited ability to analyze video frames. This method compares the statistical nature of a clear image to the statistical nature of a dusty image, assuming that the average scene's description (as characterized by an image histogram) due to surface reflectance and sun angle is invariant throughout the frame sequence. This assumption fails when shadows show up on the scene, which is evident in the last 20 s of the landing descent. In the last 20 s, the error minimization of the histogram matching is by-passed and the matching is done manually by visually comparing images.

The output of the HMM algorithm is a modified image, where "dust" has been added (mode 1) or removed (mode 0). The output image pixel $p'_{ij}$ is computed from the input image pixel $p_{ij}$:

$$p'_{ij} = \begin{cases} p_{ij} e^{-2\alpha x} + p_0 \left(1 - e^{-\alpha x}\right) & \text{mode} = 1 \\ p_0 e^{\alpha x} - \left(p_0 - p_{ij}\right) e^{2\alpha x} & \text{mode} = 0 \end{cases}. \quad \text{(C-1)}$$

where $\alpha$ is the optical extinction factor of the dust and $p_0$ is a fitting parameter associated with the dynamic range of the image (ideally $p_0 = 255$ for an 8-bit image). The factor of two in the extinction term is the result of light reflecting off of the lunar surface back to the camera. The exponent term without the factor of two corresponds to light scattered back to the camera from the dust cloud. The distance $x$ in Equation (C-1) is the effective optical-dust path length model along the camera view ray through the dust cloud, corresponding to each $ij$ image pixel in the image. It is equal to the physical path length $x_D$ of the dust for $r \leq a_0$, where $r$ is the radial distance from the engine nozzle centerline and $a_0$ is a parameter. For $r > a_0$, the effective path length is $x_D$ scaled by the radial dispersion factor:

$$x = \begin{cases} x_D & r \leq a_0 \\ x_D \left(\dfrac{a_0}{r}\right)^2 & r > a_0 \end{cases}, \tag{C-2}$$

$$x_D = \frac{\tan\theta \cot\phi \left(l_c + h\tan\phi\right) \sec\left(\phi + \tan^{-1}(q\,d/F)\right)}{\cot\left(\phi + \tan^{-1}(q\,d/F)\right) + \tan\theta}, \tag{C-3}$$

where $F$ is the focal length of the camera ($F = 10$ mm), $d$ is the pixel width ($d \approx 15$ $\mu$m), $\theta$ is the dust angle relative to horizontal ($\theta \approx 3°$), $\phi$ is the camera angle relative to vertical ($\phi = 33°$), $l_c = 1.2$ m is the camera offset distance from the nozzle center line, and $h$ is height of the LM above the surface. The variable $q$ is the vertical distance in the image in pixel units from the $ij$ pixel to a horizontal centerline in the rotated camera view:

$$q = \left(j - \tfrac{1}{2}N\right)\cos\zeta - \left(i - \tfrac{1}{2}M\right)\sin\zeta, \tag{C-4}$$

where $\zeta$ is the camera rotation angle about the camera axis ($\zeta \approx -33°$), $i$ is the horizontal pixel index, $j$ is the vertical pixel index, $N$ is the total number of horizontal pixels, and $M$ is the total number of vertical pixels.

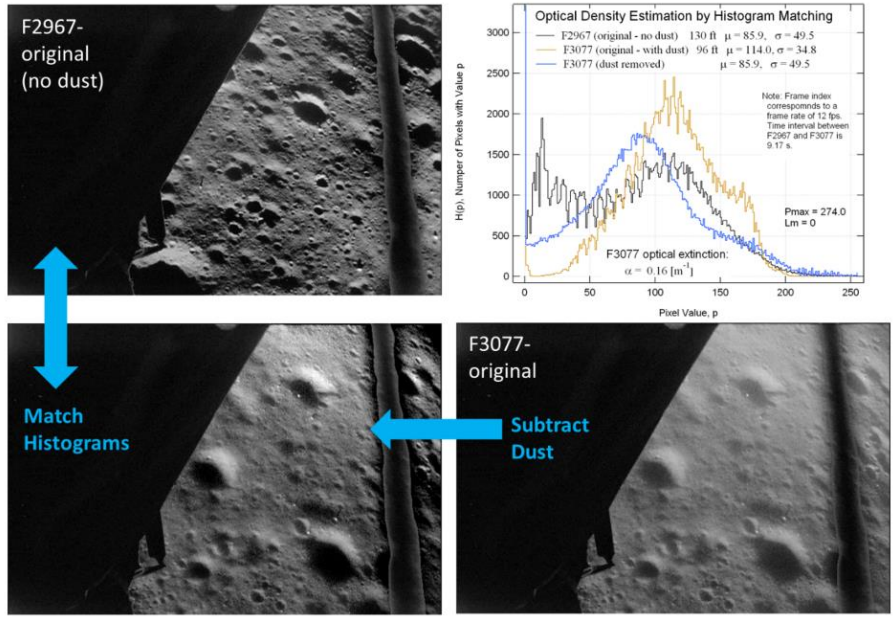

Fig. 18. HMM algorithm example for mode = 0: (lower right) image $p_{ij}$; (lower left) output image $p'_{ij}$; (upper left) reference image $q_{ij}$; (upper right) histograms for the three images.

The left side of Figure 16 shows frame F3077 ($h = 34$ m) of the cockpit video camera. The right side of Figure 16 displays a map of the same field of view for this frame, showing the effective optical-dust path length model, Equation (C-2), which is based on the tilt of the camera and increasing depth of the dust layer towards the top of the image. The horizontal and vertical axes of the plot are in pixel units. The contours are graded in increments of 0.3 m, starting with the minimum, $x = 0.3$ m at the bottom (purple in the on-line version), to a maximum, $x = 3.0$ m at the top (red in the on-line version).

Figure 17 is a similar image set, occurring 38.8 s later at an LM altitude of $h = 11$ m. The contours are graded in increments of 0.017 m, starting with the minimum, $x = 0.1$ m at the bottom (purple in the on-line version), to a maximum, $x = 0.25$ m near the center (red in the on-line version). Note that the video frame numbers F3077 (Figure 16) and F3543 (Figure 17) correspond to a constant frame rate of 12 fps.

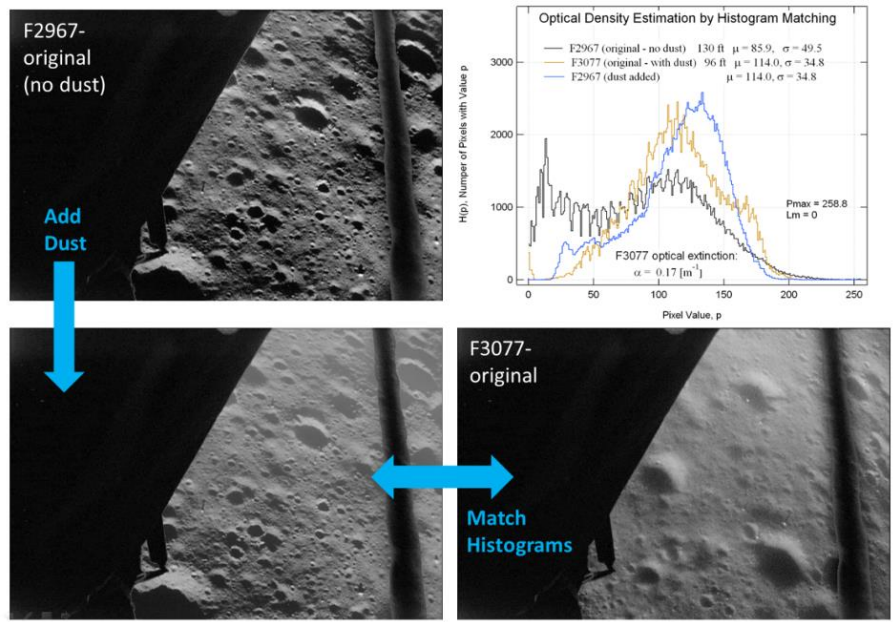

Fig. 19. HMM algorithm example for mode = 1: (upper left) input image $p_{ij}$; (lower left) output image $p'_{ij}$; (lower right) reference image $q_{ij}$ ; (upper right) histograms.

The HMM algorithm processes two input images, pixel by pixel. The first input image is represented by pixel $p_{ij}$ as shown by Equation (C-1). A reference image is represented by $q_{ij}$. For mode = 0, The HHM algorithm applies the transformation described by Equation (C-1) to the input image $p_{ij}$ (frame with dust), creating an output image $p'_{ij}$ (artificially removed dust), as shown in Figure 18. The reference image $q_{ij}$ (no dust) is then compared to $p'_{ij}$ and by matching the average and standard deviation of the their histograms, the parameters $\alpha$, $p_0$, and $a_0$ are found. Figure 19 shows a similar example for mode = 1.

R e f e r e n c e s